\documentclass{emulateapj}
\usepackage{graphicx}
\usepackage{amssymb}
\usepackage{rotating}
\usepackage{amsmath,wasysym}
\usepackage{natbib}
\usepackage{hyperref}
\begin{document}

\title{Explaining the coexistence of large-scale and small-scale magnetic fields in fully convective stars}

\author{Rakesh K. Yadav\altaffilmark{1,2}, Ulrich R. Christensen\altaffilmark{2}, Julien Morin\altaffilmark{3}, Thomas Gastine\altaffilmark{2}, Ansgar Reiners\altaffilmark{4}, Katja Poppenhaeger\altaffilmark{1}, Scott J, Wolk\altaffilmark{1}}
\affil{$^1$Harvard-Smithsonian Center for Astrophysics, 60 Garden Street, 02138 Cambridge, USA}
\affil{$^2$Max-Planck-Institut f\"ur Sonnensystemforschung, Justus-von-Liebig-Weg 3, 37077 G\"{o}ttingen, Germany}
\affil{$^3$LUPM, Universit\'e de Montpellier, CNRS, Place Eug\`ene Bataillon, 34095, France}
\affil{$^4$Institut f\"{u}r Astrophysik, Universit\"{a}t G\"{o}ttingen, Friedrich Hund Platz 1, 37077 G\"{o}ttingen, Germany}

\email{rakesh.yadav@cfa.harvard.edu}

\begin{abstract}
Despite the lack of a shear-rich tachocline region low-mass fully convective stars are capable of generating strong magnetic fields, indicating that a dynamo mechanism fundamentally different from the solar dynamo is at work in these objects. We present a self-consistent three dimensional model of magnetic field generation in low-mass fully convective stars. The model utilizes the anelastic magnetohydrodynamic equations to simulate compressible convection in a rotating sphere. A distributed dynamo working in the model spontaneously produces a dipole-dominated surface magnetic field of the observed strength. The interaction of this field with the turbulent convection in outer layers shreds it, producing small-scale fields that carry most of the magnetic flux. The Zeeman-Doppler-Imaging technique applied to synthetic spectropolarimetric data based on our model recovers most of the large-scale field. Our model simultaneously reproduces the morphology and magnitude of the large-scale field as well as the magnitude of the small-scale field observed on low-mass fully convective stars.
\end{abstract}

              
%

\section{Introduction}

The tachocline region inside the Sun, where the rigidly rotating radiative core meets the differentially rotating convection zone, is thought to be crucial for generating the Sun's magnetic field~\citep{charbonneau2005}. Low-mass fully convective stars do not possess a tachocline and were originally expected to generate only weak small-scale magnetic fields~\citep{durney1993}. Observations, however, have painted a different picture of magnetism in rapidly-rotating low-mass fully convective stars: (1) Zeeman broadening measurements revealed average surface field of several kiloGauss~\citep{krull1996, krull2000, reiners2009}; on the Sun such field strength is found only in the sunspots~\citep[e.g.][]{solanki2003}. (2) Zeeman-Doppler-Imaging (ZDI) technique discovered fields with a morphology often similar to the Earth's dipole-dominated field~\citep{donati2006, morin2008, morin2010}. (3) Comparison of Zeeman broadening and ZDI results showed that more than 80\% of the magnetic flux on the these stars resides at small scales~\citep{reiners2009b}.

So far, theoretical and computer simulation efforts have not been able to reproduce these features simultaneously. An earlier fully convective dynamo simulation study~\citep{dobler2006} produced dominantly dipolar fields for a low density contrast of about 10 in the convection zone, while the dipolar mode was weak in another study~\citep{browning2008} with density contrast of about 100. The contribution from small-scale fields was rather small in the former and too large in the latter. Systematic numerical simulations~\citep{gastine2012b} showed that dipole-dominant field morphology is progressively destabilized as the density stratification in the simulated convection zone increases. In highly stratified simulations magnetic field generation shifts to the outer layers where the flow is more vigorous and has the highest magnetic Reynolds number $Rm=u\,D/\lambda$ (ratio of magnetic induction and magnetic diffusion; $u$ is local mean velocity, $D$ is thickness of convection zone, and $\lambda$ is the magnetic diffusivity). On the other hand, modeling approaches based on mean-field dynamo theory (which parameterizes effects of small-scale fields) has only produced non-dipole dominant fields in fully convective (FC) star models~\citep{kuker1999, chabrier2006, shulyak2015}.

The rather peculiar nature of the magnetism in FC stars has prompted investigations which are not based on the canonical solar dynamo model. Remarkably, an empirical scaling law derived from dynamo simulations tailored to model the magnetism in the Earth has been able to account for the kG strength of the magnetic field in FC stars~\citep{christensen2009}, suggesting that a similar dynamo mechanism might be operating in both. Differential rotation in the Earth supposedly plays little role in sustaining the geomagnetic field (see~\citet{roberts2013} for a review). In the classical ``mean-field" formulation such dynamos are referred to as ``$\alpha^2$-dynamos" where only the helical turbulence is the main driver of dynamo activity. However, given the vastly different convection zones of the Earth and stars (i.e. nearly incompressible liquid metal in the former and highly compressible plasma in the latter) it is not clear how a similar dynamo mechanism might work in these objects. The geodynamo simulations have reproduced many observed features of the Earth's magnetic field~\citep{roberts2013}. However, modelling the FC star dynamo remains challenging. Except for the agreement between the mean field strength in FC stars and theoretical models~\citep{chabrier2006, yadav2013b, schrinner2014} no stellar dynamo model has been able to fully reproduce the detailed magnetic field structure of FC stars.

\section{Methods}
\subsection{Simulation}
We study the magnetic field generation mechanism in low-mass FC stars using direct numerical simulation of the anelastic~\citep{lantz1999} magnetohydrodynamic equations using the open-source MagIC code\footnote{Available at \href{https://github.com/magic-sph/magic}{\tt https://github.com/magic-sph/magic}}~\citep{gastine2012a}. The relevant formalism is described in detail in \citet{yadav2015}. We consider a nearly fully convective sphere and the ratio of the inner ($r_i$) and the outer ($r_o$) boundary radius is 0.1. The very small stagnant central region (0.1\% by volume)  is retained for practical reasons. The density varies by a factor $\approx$150 (five density scale heights) in the convection zone. The fluid is assumed to be an ideal gas with a polytropic index of 1.5. The various non-dimensional control parameters are defined as follows: the Prandtl number $P_r=\nu/\kappa$, where $\nu$ is the viscosity and $\kappa$ is the thermal diffusivity; the magnetic Prandtl number $P_m=\nu/\lambda$, where $\lambda$ is the magnetic diffusivity; the Ekman number $E=\nu(\Omega_s D^2)^{-1}$, where $\Omega_s$ is the rotation rate of the shell and $D=r_o-r_i$ is the shell thickness; the Rayleigh number $Ra=g_{o}\,D^{3}\,\Delta s (c_{p}\,\nu\,\kappa)^{-1}$, where $g_o$ is the gravity at the outer boundary, $\Delta s$ is the fixed entropy contrast between the boundaries, and $c_p$ is the specific heat at constant pressure. The gravity varies linearly with radius.

To model rotationally dominated convection we set the Ekman number to $10^{-5}$ (or a Taylor number of about $10^{10}$), smaller than the values used in earlier studies~\citep{dobler2006,browning2008, gastine2012b}. We chose a relatively low Prandtl number of 0.1, which helps to distribute the convection throughout the shell and may contribute to sustaining a strong-field dynamo action in the deep interior of the shell~\citep{jones2014}. The magnetic Prandtl number is set to 2 to attain high $Rm$ values throughout the convection zone. The Rayleigh numbers is set to $3\times10^8$. The various diffusivities are considered constant throughout the spherical shell. Both boundaries are stress-free and insulating.

A grid resolution of (100, 320, 640), where the three numerals indicate number of grid points in radial, meridional, and azimuthal direction, respectively, was used for most part of the simulation. The final stages were performed at a much higher resolution of (160, 1024, 2048) to check the stability of the results. The simulation was evolved for about 7500 rotations (about a quarter of the magnetic diffusion time), consuming $\approx$0.8 Million CPU hours. The entire simulation data is available upon request to the authors. Note that a central source of energy assumed in our model is more consistent with a low-mass fully convective star than a fully convective pre-main-sequence T Tauri star where the gravitational contraction provides an energy source distributed throughout the convection zone~\citep[e.g. see][]{bessolaz2011}.

\begin{figure*}
\epsscale{0.48} \plotone{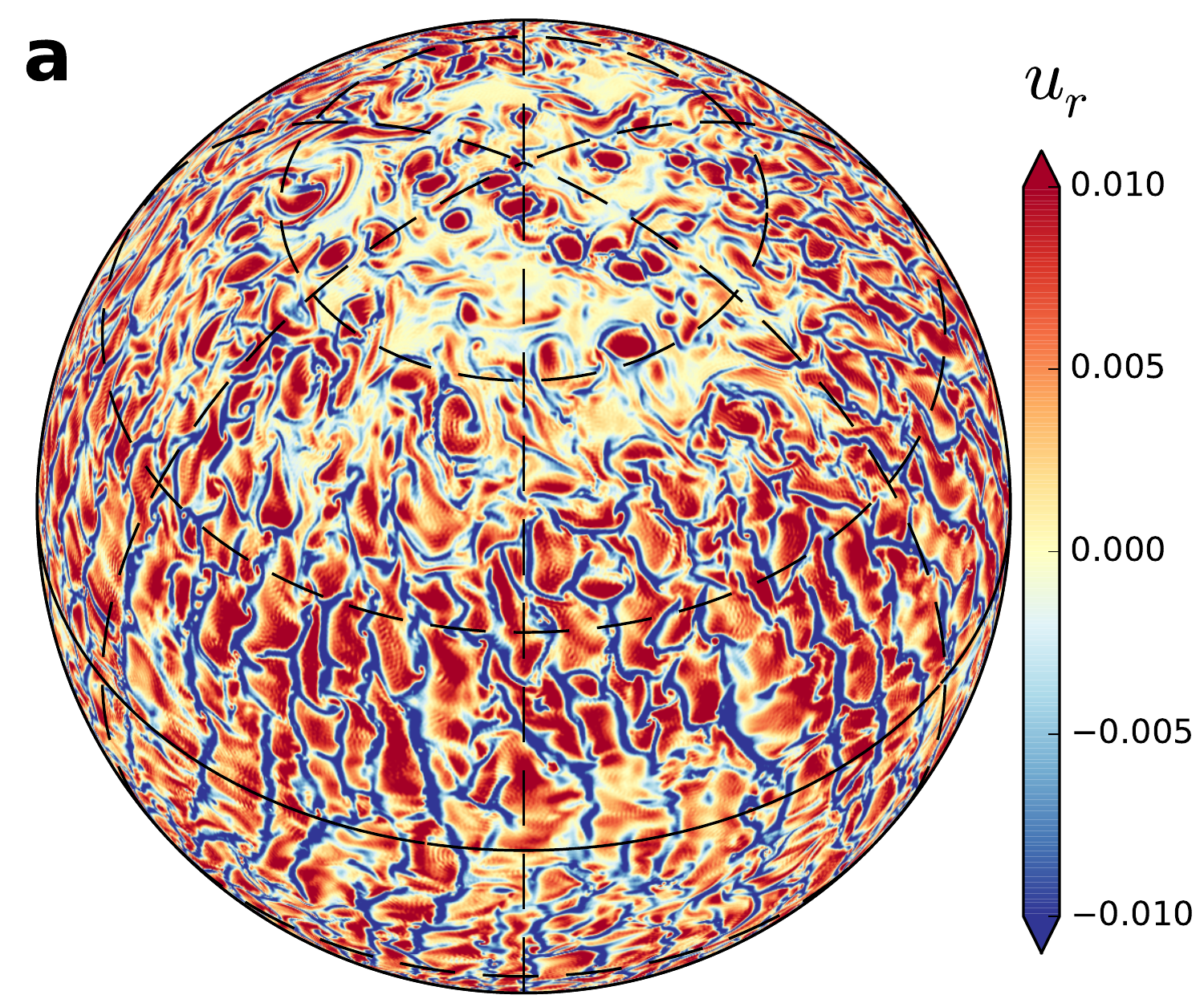} \epsscale{0.48}\plotone{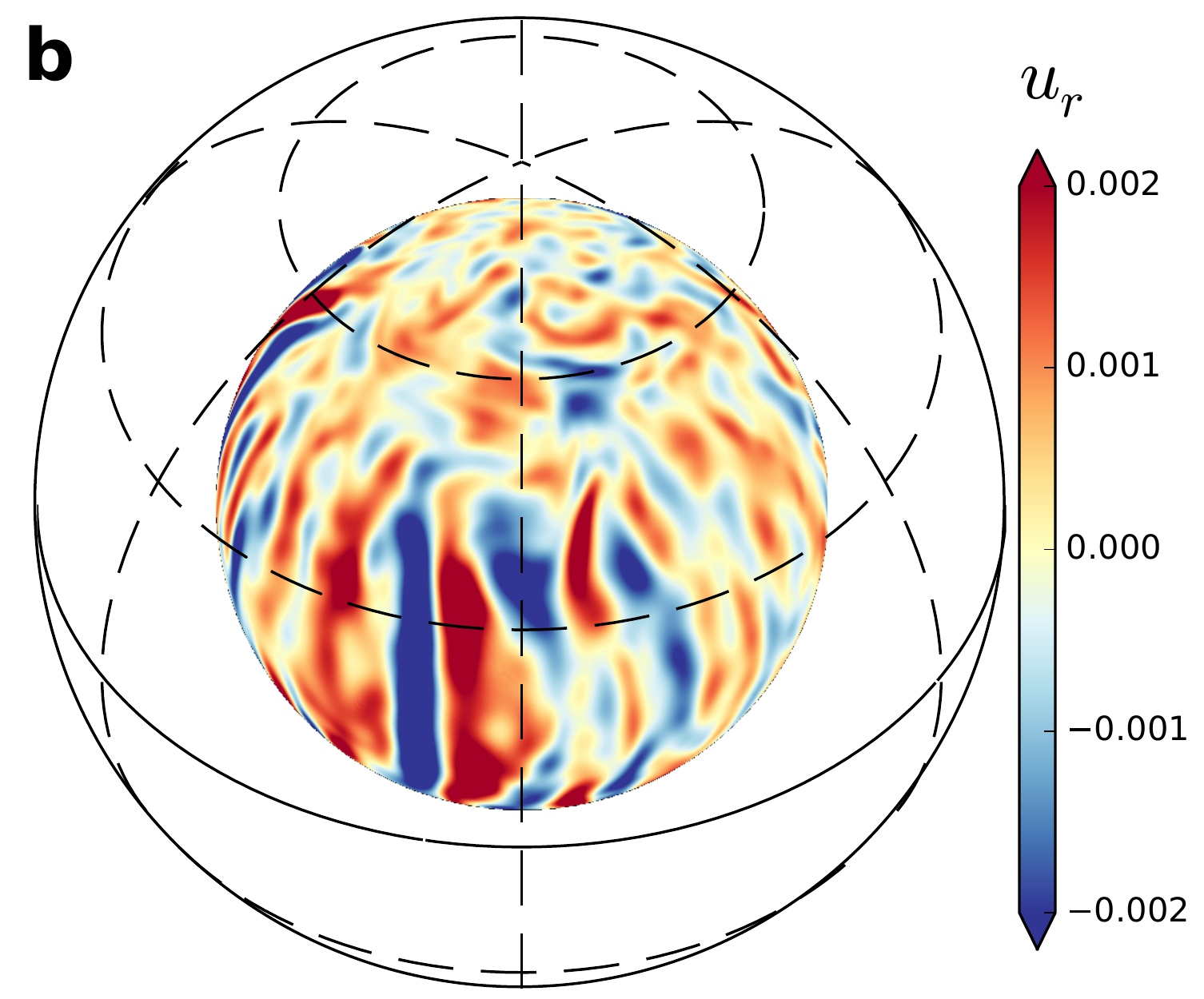} \\
\epsscale{0.48} \plotone{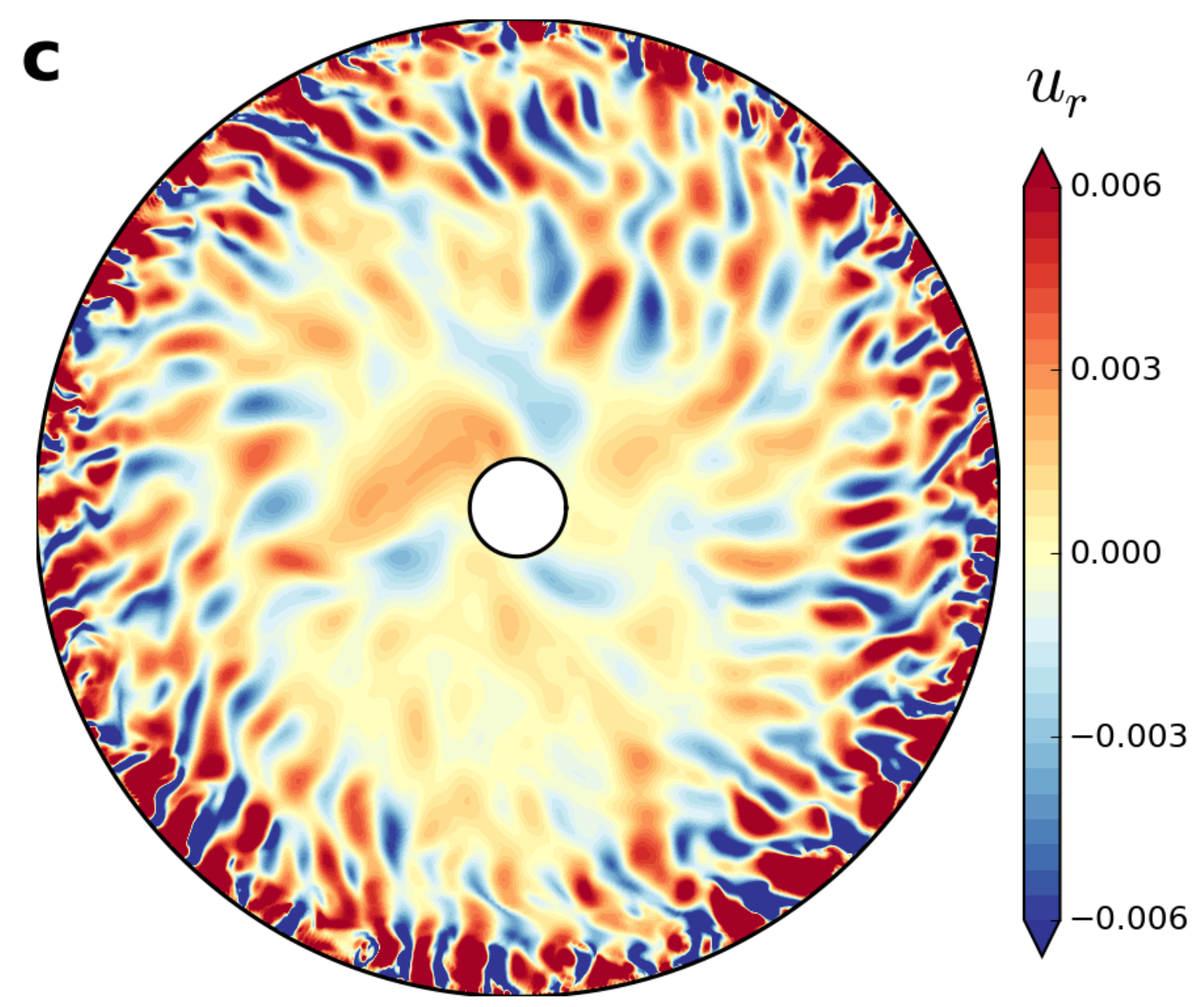} \epsscale{0.48}\plotone{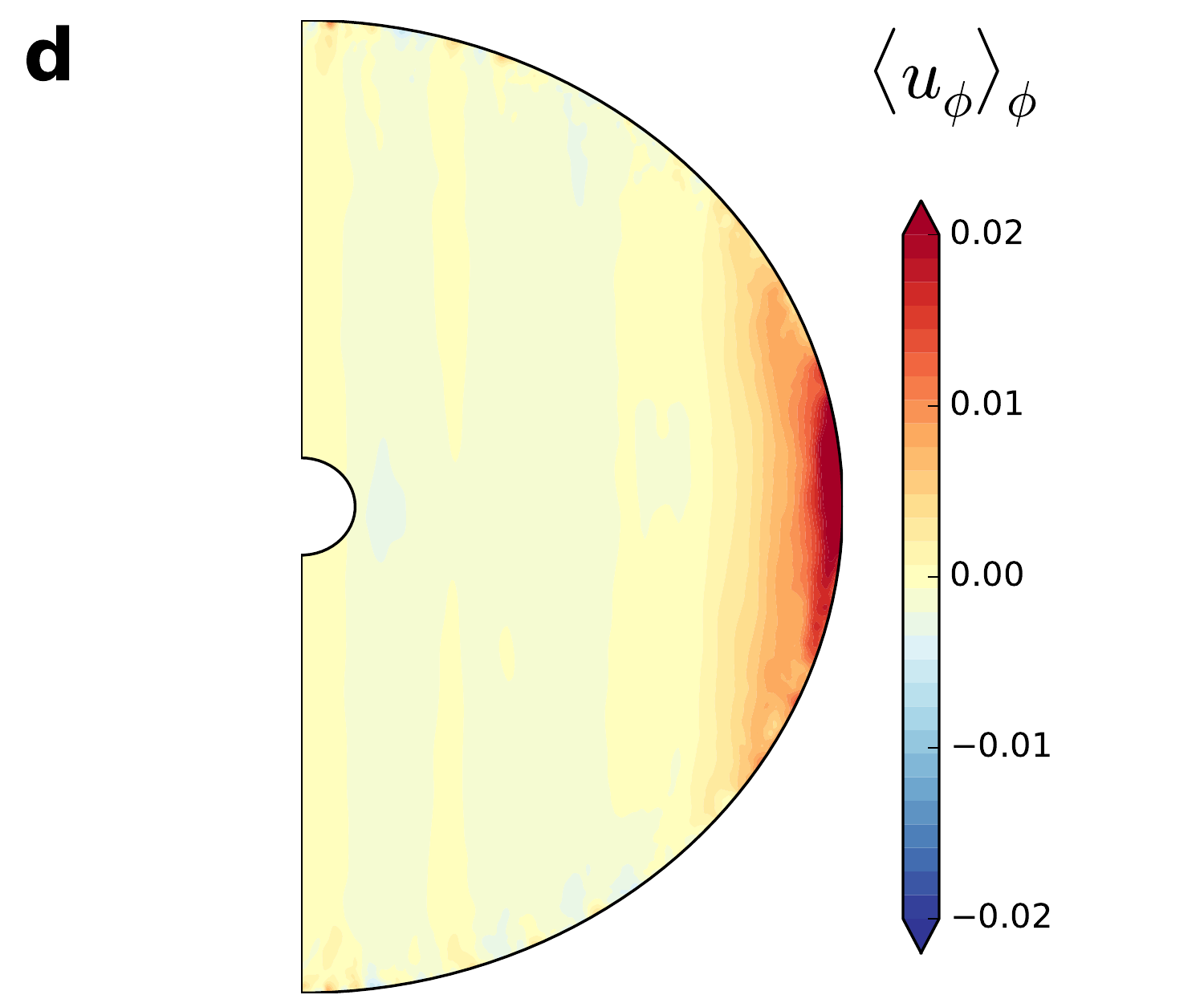}
\caption{Orthographic projections of the radial velocity $u_r$ at a radius of $0.99r_o$ in panel {\bf a} and at $0.6r_o$ in panel {\bf b}. The viewing angle is $45^{\circ}$ north of the equator and the wire-frame represents the simulation outer boundary. Radial velocity on the equatorial plane of the simulation in panel {\bf c}. Azimuthally averaged azimuthal velocity $\langle u_{\phi}\rangle_{\phi}$ on a meridional plane of the simulation in panel {\bf d}. All the quantities are represented in terms of the Rossby number. The color scale is saturated at values lower than the extrema to highlight fainter structures. \label{Fig1}}
\end{figure*}

\subsection{Scaling to physical units}
The non-dimensional model can, in principle, be applied to different types of fully convective stars. To express results in an exemplary way in physical units, we have scaled them, using the results of a stellar evolution model~\citep{granzer2000}, to a main sequence star with 30\% of the solar mass and an effective surface temperature of 3680 K. For scaling we took the values for radii, mean density, thermodynamic parameters and gravity from the stellar model. Then we are left with two choices. We could either set the rotation rate to a typical value of a rapidly rotating M-dwarf, say 1 day, or we could fix the energy flux (luminosity) to the correct value. Most stellar convection modelers chose the first option, however, with the consequence that the luminosity in their model must be larger than actual values by several orders of magnitude~\citep{dobler2006}. Here we take the second option because we are mainly interested in realistic values of the magnetic field strength, which is primarily controlled by the luminosity~\citep{christensen2009} in the class of dynamos we are interested in. Values for the diffusivities and the rotation rate are then determined in order to match the values of the nondimensional model parameters. The diffusivities are $\nu$ = 10$^6$ m$^2$~s$^{-1}$, $\lambda$ = 0.5$\times$10$^6$ m$^2$~s$^{-1}$, $\kappa$ = 10$^7$ m$^2$~s$^{-1}$, i.e. much higher than molecular values and must be understood as effective turbulent diffusivities. The outermost radius of the model ($r_o$=197,550 km) would be at 95\% of the radius of the stellar model on the basis of a radial density contrast of $\approx$150. The density and gravity at $r_o$ are 617 kg m$^{-3}$ and 1000 m s$^{-2}$, respectively. The luminosity is 0.0147 times the solar value. The rotation period is about 20 days. From the point of view of ZDI this rotation rate is not rapid enough to construct a reliable magnetic field map. However, from the perspective of magnetic activity saturation in the type of star considered here 20 days rotation period is close to the $\approx$15 days rotation period where the activity starts saturating~\citep[see Eqn.~10 in][]{reiners2014}.

\subsection{Applying ZDI to simulation}
Zeeman-Doppler-Imaging is a tomographic imaging technique aimed at mapping the large-scale component of stellar magnetic fields from spectropolarimetric time-series~\citep{semel1989}. For cool stars ZDI reconstructions are generally based on Stokes~$V$ (net circular polarisation) average line profiles computed with the Least Square Deconvolution (LSD) method~\citep{donati1997, kochukhov2010}. We use the magnetic field on the surface of our simulation to generate a synthetic times-series of Stokes~$V$ line profiles using the forward module of ZDI. These synthetic datasets have been computed for line parameters typical of LSD line profiles of M dwarfs (central wavelength $\lambda_0=700$~nm, Land\'{e} factor 1.2) and according to the specifications of the ESPaDOnS and NARVAL spectropolarimeters~\citep{donati2006b} (spectral resolution $R$  = 65,000 and spectral sampling $\Delta v=1.8$ km s$^{-1}$). The resulting datasets are typical of those obtained in observational studies~\citep{morin2008}. Gaussian noise is added to each line profile to obtain a signal to noise ratio of 5,000 (or $\sigma_V=2\times10^{-4} I_c$, where $I_c$ is the unpolarised continuum level).

In our ZDI implementation, the magnetic field distribution is reconstructed from these synthetic datasets directly as a set of coefficients of a poloidal-toroidal decomposition projected on a spherical harmonics basis~\citep{donati2006c}. It should be noted that the performed ZDI reconstructions are idealized in the sense that real datasets never achieve perfect even sampling of the stellar rotation cycle, and the line model used to describe Stokes~$V$ line profiles is exactly the one used to generated them. Nonetheless, the synthetic ZDI maps are representative of the capabilities of the technique and of its intrinsic imaging limitations.

\section{Results}
A non-magnetic simulation with the control parameters mentioned above generates very strong radial and latitudinal differential rotation. The simulation undergoes dramatic changes after a weak seed magnetic field is introduced. The magnetic field grows exponentially until its energy is in rough equipartition with the kinetic energy. In the saturated state, the Rossby number $Ro$ = $u(\Omega_s\,D)^{-1}$ ($\Omega_s$ is the bulk rotation rate) varies from about 0.003 to about 0.04 in the convection zone. $Ro$ measures the ratio of inertial forces, which promote turbulence, to Coriolis forces, which have an ordering effect on the flow. $Rm$ varies from about 550 in the deep interior to about 9,000 in the shallow layers. The ratio of total heat transported to the heat diffused from bottom to the top boundary, called the Nusselt number $Nu$, is $\approx$1.3. Due to the variation in $Ro$ chaotic convection at shallow depth (Fig.~\ref{Fig1}a) coexists with axially-aligned columnar convection (Fig.~\ref{Fig1}b) in the deeper layers~\citep{browning2008, gastine2012b, hotta2015a}. The length scale of convection also changes with the radius due to the high density stratification in the convection zone (Fig.~\ref{Fig1}c). Differential rotation is severely quenched due to the Lorentz forces associated with the magnetic field. Significant azimuthal flows exist near the outer surface only at low latitudes (Fig.~\ref{Fig1}d), here the relative differential rotation $\Delta\Omega/\Omega_s$ (where $\Delta\Omega$ is the difference in equatorial and  polar rotation rate) is about 0.02 which translates to an equator-to-pole lap time of about 50 rotations. Such a small value of differential rotation is consistent with  observations~\citep{morin2008, reinhold2013, davenport2015} and earlier numerical studies~\citep{browning2008} of FC stars.

\begin{figure}
\epsscale{1.2} \plotone{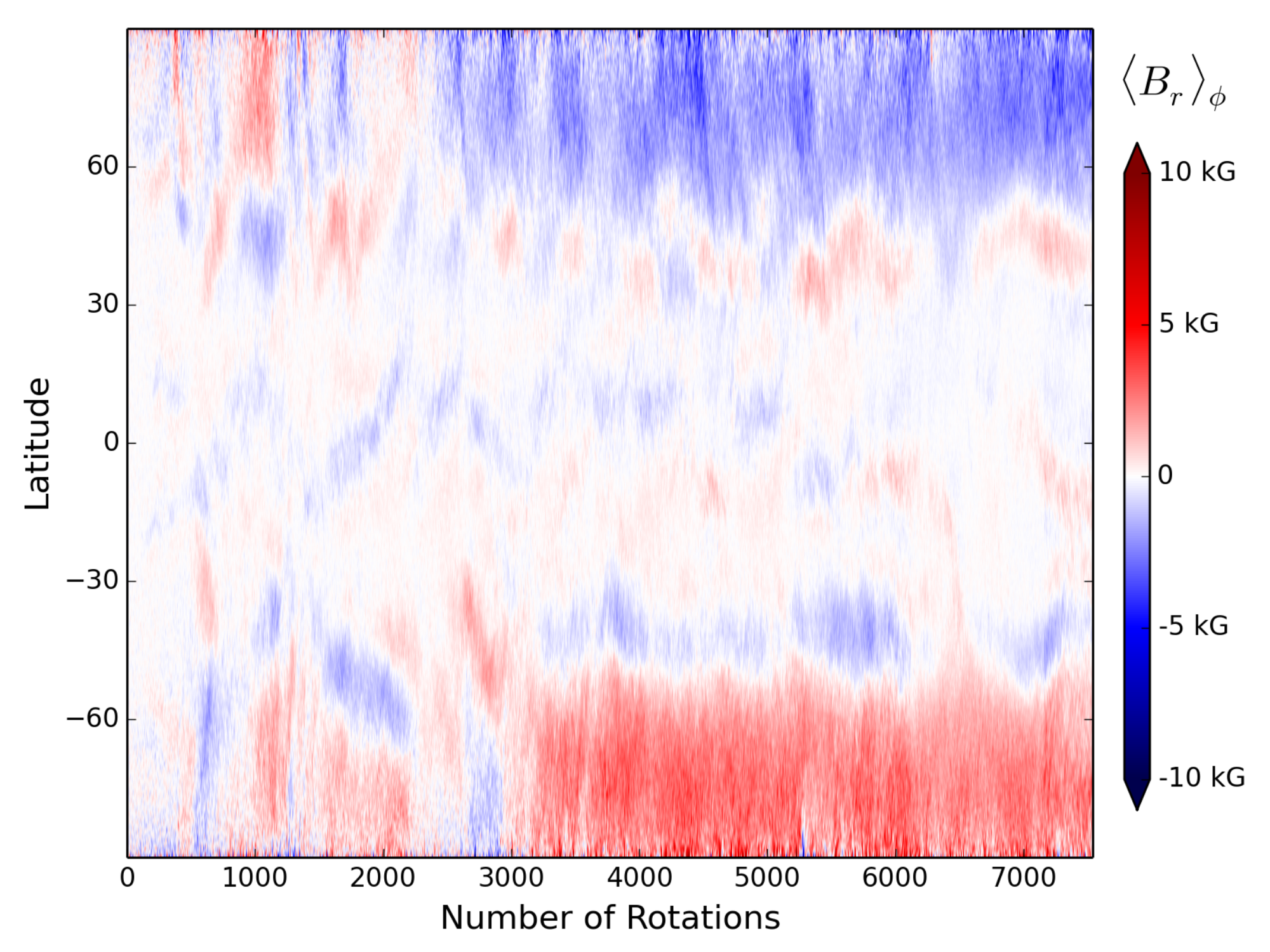} 
\caption{Temporal evolution of the azimuthally averaged radial component of the magnetic field at a radius of $0.99r_o$. \label{Fig2}}
\end{figure}

\begin{figure}
\epsscale{1.15} \plotone{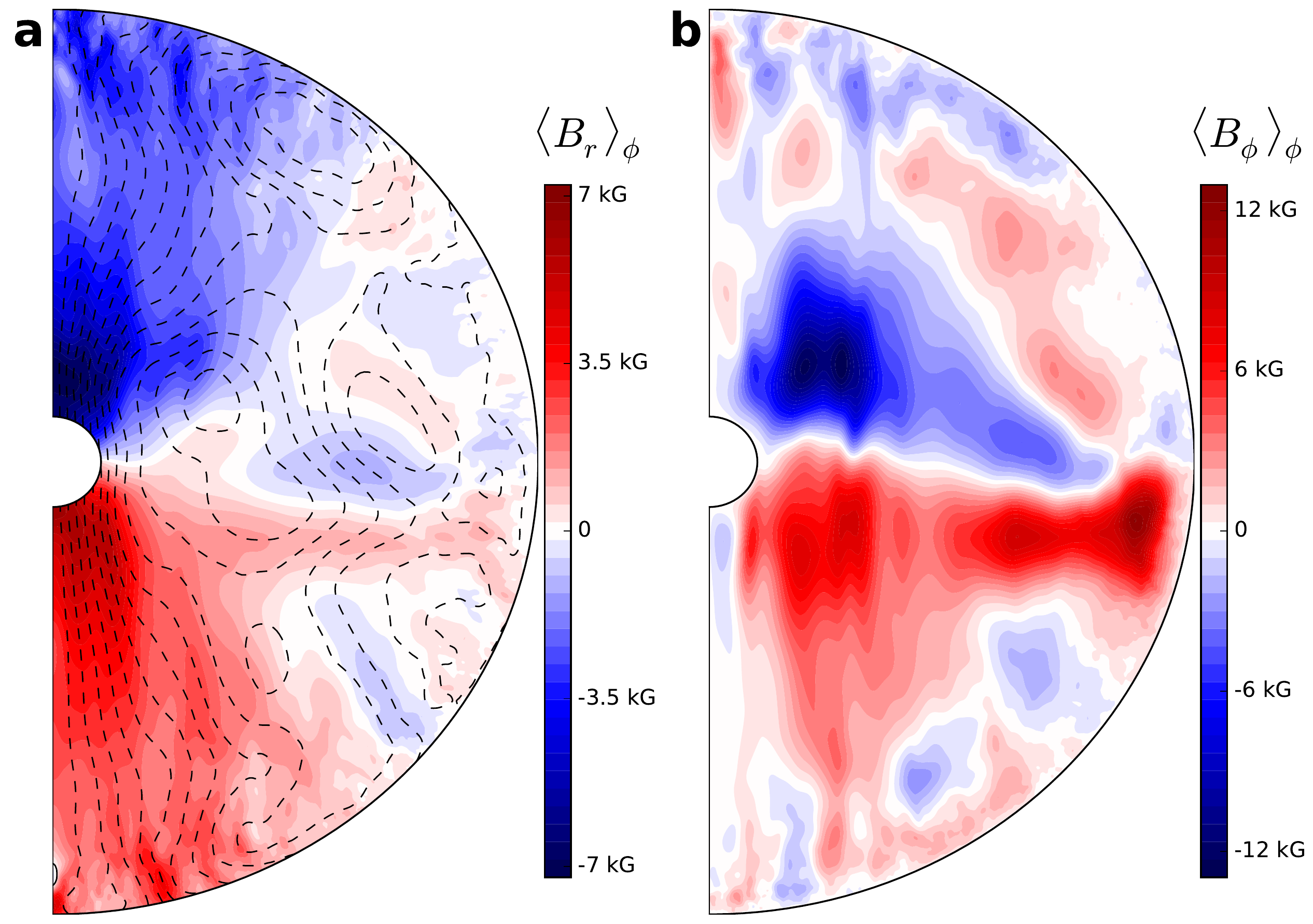} 
\caption{Azimuthally averaged radial magnetic field as colored contours and azimuthally averaged poloidal magnetic field as dashed lines in panel {\bf a}. Azimuthally averaged azimuthal magnetic field in panel {\bf b}. The figure is a snapshot in time. \label{Fig3}}
\end{figure}

\begin{figure}
\epsscale{1.15} \plotone{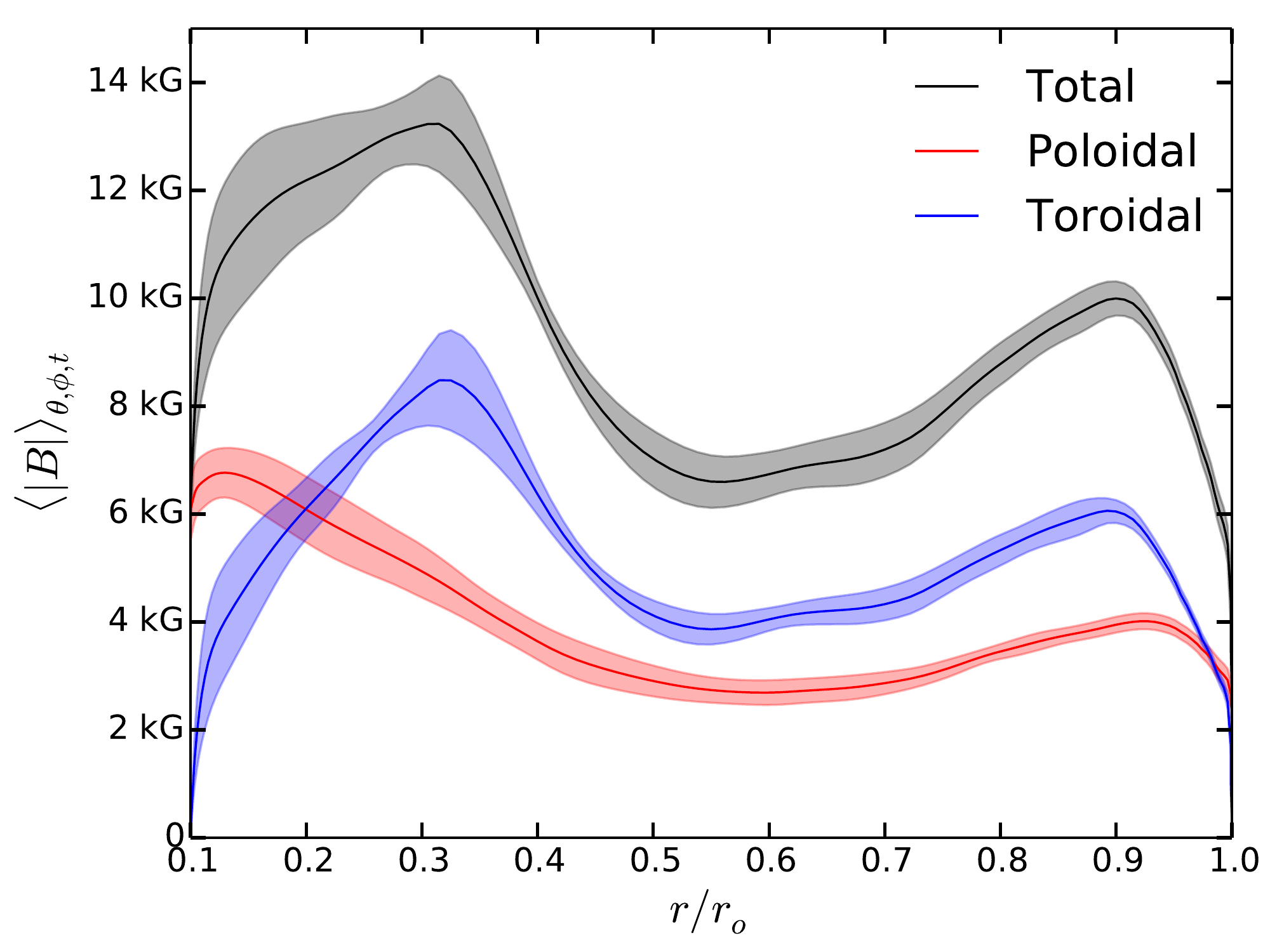} 
\caption{Radial variation of the unsigned mean poloidal, toroidal, and total magnetic field. The values are averaged latitudinally, azimuthally, and temporally over $\sim$200 rotations. The shaded region around each curve highlights the standard deviation around the mean. \label{Fig4}}
\end{figure}

A saturated state is reached in the simulation after  about 3000 rotations (Fig.~\ref{Fig2}). In this state, the large-scale field morphology is dominated by an axisymmetric dipolar mode (Fig.~\ref{Fig3}a) that is stable over long time scales. Both the poloidal and toroidal magnetic field intensity peak in the interior (Fig.~\ref{Fig3} and Fig.~\ref{Fig4}), suggesting that a strong $\alpha^2$-type dynamo operates on the deep-seated convection where shear is almost absent. Note that a peak field intensity of about 14 kG in the simulation interior (Fig.~\ref{Fig4}) does not support the assumption of mega Gauss field strength considered in some interior models of fully convective stars~\citep{feiden2014}.

\begin{figure*}
\epsscale{0.3} \plotone{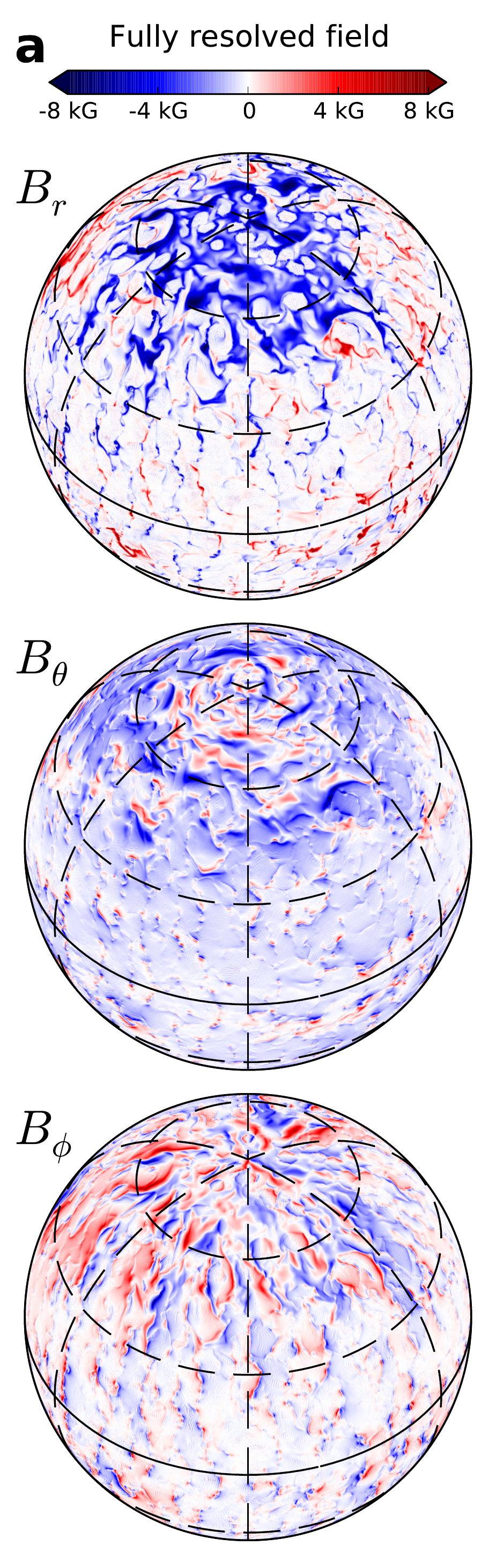} \epsscale{0.3}\plotone{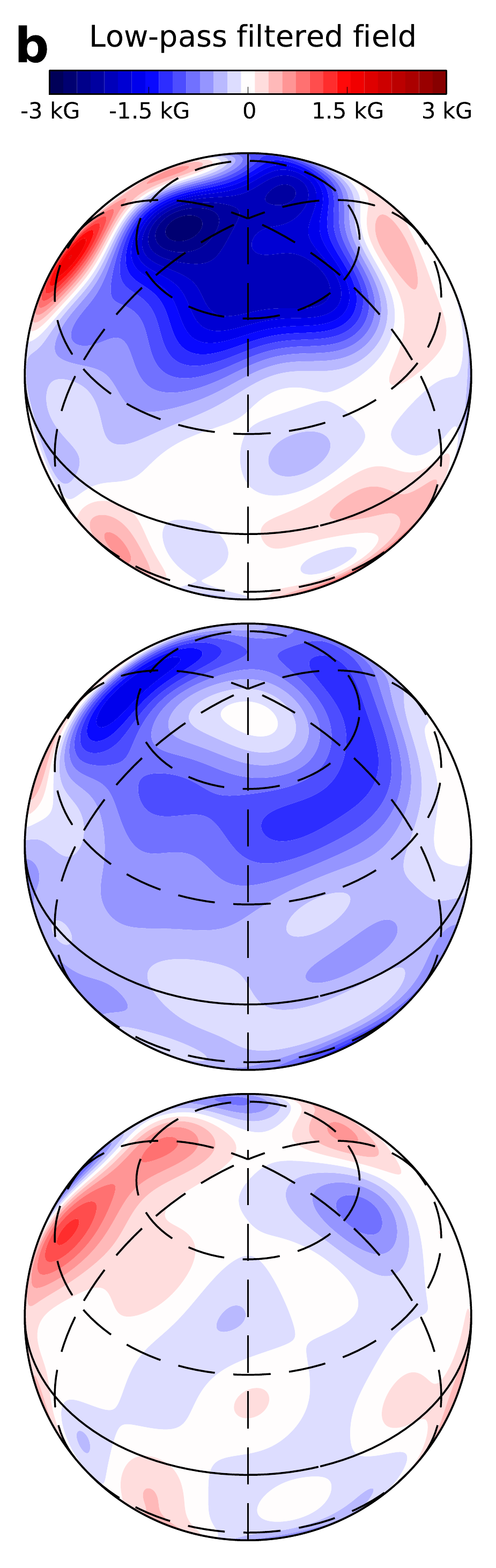} 
\epsscale{0.3} \plotone{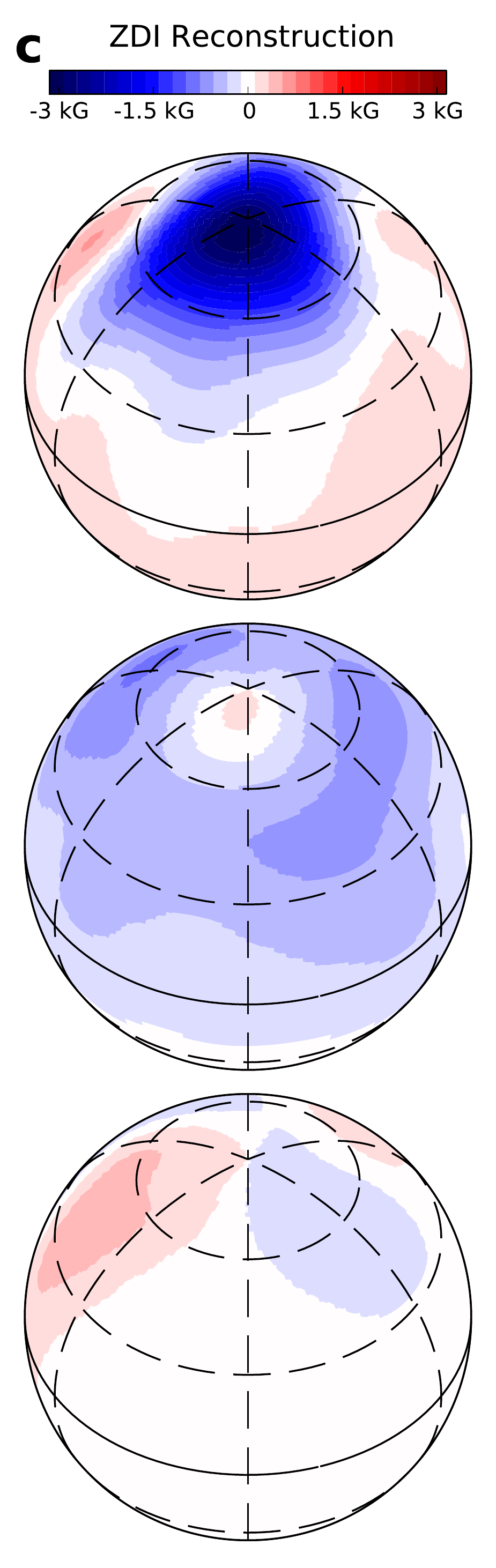} 
\caption{Orthographic projections of the radial $B_r$, meridional $B_\theta$, and azimuthal $B_\phi$ component of the magnetic field at the outer boundary in panel {\bf a}. Panel {\bf b} shows orthographic projections resulting after applying a low-pass filter (harmonic degree up to 10) to the data in panel {\bf a}. Magnetic field maps constructed by applying the ZDI technique to the simulation data in panel {\bf a} are shown in panel {\bf c}. A $v\sin i$ of 20 km s$^{-1}$ was used for constructing the ZDI maps. \label{Fig5}}
\end{figure*}

The more chaotic convection in the outer layers is shredding the magnetic field at the surface (Fig.~\ref{Fig5}a), where magnetic flux is concentrated into narrow downwellings (also known as ``inter-granular lanes"). Such concentration of magnetic field is a generic feature of compressible convection~\citep{stein2012, beeck2015}. However, at large scales, the strong dipolar component survives this pummeling by the convection and provides a diffuse carpet of strong magnetic flux with a single dominant polarity at high latitudes in each hemisphere (Fig.~\ref{Fig5}a). The strong field suppresses near-surface convection in rather extensive regions at high latitudes (Fig.~\ref{Fig1}a), reminiscent of polar dark spots recently reported in a few fully convective stars~\citep{barnes2015}. Similar suppression of convection by magnetic field was also observed in a recent dynamo simulation of a K-type star~\citep{yadav2015}. In passing we note that a small-scale dynamo~\citep{cattaneo1999} may be present at shallow depth in our model due to the vigorous chaotic convection. However, we have not yet investigated its presence in our model.

Remarkably, bipolar magnetic field structures are present nearly all over the surface (Fig.~\ref{Fig5}a). Such regions can provide the necessary twisted magnetic field lines which generate X-ray flares~\citep{haisch1991}, and may explain the very high level of X-ray flare activity of FC stars~\citep{moffett1974}. Furthermore, a recent analysis of flares on several FC stars found almost no correlation between the stellar phase and flare occurrence rate~\citep{hawley2014}. The authors posit that either a large polar spot group (presumably the main source of flares) is closely aligned with the rotation axis or that the whole surface is covered with bipolar active regions. Our simulation favours the latter scenario.

The observational technique based on the Zeeman broadening is sensitive to the unsigned mean magnetic field present at all scales on the stellar surface. Studies based on this principle~\citep{krull1996, krull2000, reiners2009} show field strengths of about 2 to 4 kG in rapidly rotating FC stars. The surface averaged value of the magnetic field at the outer boundary is $\approx$2.3 kG in our model (Fig.~\ref{Fig5}a). On the other hand, Zeeman-Doppler-Imaging utilizes the disc-integrated circular polarization in atomic spectral lines to construct a map of the vector magnetic field of stars~\cite{semel1989}. However, it is prone to cancellation effects, and can reveal only the large scale magnetic field structures. This technique has been applied to several rapidly rotating FC stars and has frequently revealed~\citep{morin2008} a dipole-dominated magnetic field with a surface averaged strength of roughly 600 Gauss. We have applied this same technique to synthetic spectra which we calculated on the basis of the surface magnetic field of our simulation to explore how our artificial star would look like to an observer. The synthetic maps (Fig.~\ref{Fig5}c) reconstruct the large scale features present in the simulation (Fig.~\ref{Fig5}b) to a good extent, however, they underpredict the field amplitude~\citep{rosen2015}. For instance, about 450 G surface-averaged field is recovered in ZDI maps (Fig.~\ref{Fig5}c) as compared to about 1.1 kG at large scales (up to harmonic degree 10; Fig.~\ref{Fig5}b) and about 2.3 kG at all scales present at the simulation surface (Fig.~\ref{Fig5}a). The recovery fraction of about 20\% of the total flux in the synthetic ZDI maps is consistent with the observations of the FC star magnetic field where ZDI recovers only about 14-20\% of the total magnetic flux observed via the Zeeman broadening measurements~\citep{reiners2009b}.  We also performed the ZDI reconstruction with different inclination of the stellar rotation axis to the line-of-sight ($i=20^{\circ},70^{\circ}$) and of the projected equatorial rotation velocity of the star ($v\sin i=10, 40$ km s$^{-1}$). We recovered similar results with higher/slower $v\sin i$ values producing finer/smoother magnetic field structures.

\section{Summary and Prospects}
We presented a fully non-linear global dynamo model which, for the first time, spontaneously produces many of the observed properties of the magnetic field in low-mass fully convective stars. An $\alpha^2$-dynamo working in the interior of the simulation maintains a dipole dominant, large scale, and strong magnetic field. Turbulent convection in outer layers acts to shred the magnetic field and channels magnetic energy from large scales to small scales. These two mechanisms acting in concert are important for generating the observationally consistent magnetic field features in the model. 

Rotationally dominated convection, large enough magnetic Reynolds numbers, and high density stratification appear to be crucial for supporting the two mechanisms mentioned above. For example, reducing the rotation rate by 10 destroys the dipolar morphology of the magnetic field and only small-scale fields are left\footnote{ However, note that a direct comparison with stars is not obvious here since we do not take into account the changes in the stellar physical properties due to changed rotation rate.}. Lowering the magnetic Reynolds number by four (by using lower electrical conductivity) also destabilizes the dipolar mode. In this case large scale fields are present but the peak field intensity shifts to the outer layers and the axisymmetric poloidal  field component weakens. Simulations with mild density contrast of about 20 do not posses a sufficient scale-separation to simultaneously sustain magnetic fields at both large and small scales~\citep{dobler2006, gastine2012b}.

We used the ZDI technique to analyze the large-scale magnetic field of our simulation. We consistently recovered a strong polar spot of radial field corresponding to the visible magnetic pole of the dipolar component present in our numerical simulation. The recovered magnetic field maps feature almost no toroidal component (less than 2\% of the reconstructed magnetic energy) and are mostly axisymmetric, in agreement with the large-scale component of our numerical simulation.

The simulation generates magnetic field structures which may have a correspondence at fully convective stars. The octupolar component of the magnetic field becomes prominent at times when the dipolar component temporarily weakens (Fig.~\ref{Fig2}). If such an interplay of dipolar and octupolar modes exists on low-mass fully convective stars, then it can be observed by regularly monitoring their magnetic field morphology at different epochs.

Intriguingly there are some low-mass FC stars which presumably have similar physical properties but show either a dipole-dominated field morphology or a more complex and weaker field~\citep{morin2010}. Theoretical efforts to explain this behavior resort to the idea of weak- and strong-field dynamos~\citep{morin2011} which can be accessed in numerical simulations by choosing different initial conditions for the magnetic field~\citep{simitev2009, gastine2013a}. However, without a sound justification for different magnetic field initial conditions during the early stages of a star's evolution, this theoretical idea remains speculative. One could also imagine that FC stars might be behaving similar to the Sun where the magnetic field regularly switches between a dipole-dominant morphology and a more complex and weaker one~\citep{kitchatinov2014}. In this scenario, the phenomenon of different field morphology for otherwise similar stars is an observational manifestation of the temporal evolution of the field. In our simulation, the axial dipolar mode was rather stable and maintained its dominance. The observations and simulations of the geodynamo show that the dipolar mode chaotically reverses, and during the reversal magnetic field is multipolar and relatively weak~\citep{roberts2013}. If we assume that an Earth like dynamo is working in the deep interior of our fully convective star model, then there might be a limited range of Rossby numbers where geodynamo like reversals are possible in this model as well.

\acknowledgements
We thank the anonymous referee and J.-F. Donati for a careful reading and interesting suggestions. Funding from DFG (through SFB 963/A17 and SPP 1488) and  NASA (through the {\em Chandra} grant GO4-15011X) is acknowledged. Simulations were performed at RZG and GWDG.


\end{document}